\documentclass[runningheads]{llncs}
\usepackage{amssymb}
\usepackage{amsmath}
\usepackage{booktabs}
\usepackage{url}
\usepackage[misc]{ifsym}
\usepackage{bbding}

\usepackage[colorlinks,
linkcolor=blue,
anchorcolor=blue,
citecolor=blue]{hyperref}

\makeatletter
\def\thanks#1{\protected@xdef\@thanks{\@thanks\protect\footnotetext{#1}}}
\makeatother
\usepackage[T1]{fontenc}
\usepackage{graphicx,verbatim}
\begin{document}
\title{Hybrid-View Attention Network for Clinically Significant Prostate Cancer Classification in Transrectal Ultrasound}

\titlerunning{Hybrid-View Attention for csPCa Classification in TRUS}
%
\begin{comment}  %% Removed for anonymized MICCAI 2025 submission
\author{First Author\inst{1}\orcidID{0000-1111-2222-3333} \and
Second Author\inst{2,3}\orcidID{1111-2222-3333-4444} \and
Third Author\inst{3}\orcidID{2222--3333-4444-5555}}
%
\authorrunning{F. Author et al.}
% First names are abbreviated in the running head.
% If there are more than two authors, 'et al.' is used.
%
\institute{Princeton University, Princeton NJ 08544, USA \and
Springer Heidelberg, Tiergartenstr. 17, 69121 Heidelberg, Germany
\email{lncs@springer.com}\\
\url{http://www.springer.com/gp/computer-science/lncs} \and
ABC Institute, Rupert-Karls-University Heidelberg, Heidelberg, Germany\\
\email{\{abc,lncs\}@uni-heidelberg.de}}

\end{comment}

\author{
Zetian Feng\inst{1,\dagger}\thanks{$\dagger$ Zetian Feng and Juan Fu contribute equally to this work.} \and 
Juan Fu\inst{2,\dagger} \and
Xuebin Zou\inst{2} \and
Hongsheng Ye\inst{2}\and
Hong Wu\inst{1}\and \\
Jianhua Zhou\inst{2,}\Envelope 
\and Yi Wang\inst{1,}\Envelope\thanks{\Envelope Corresponding authors: Jianhua Zhou and Yi Wang.}
}  %% Added for anonymized MICCAI 2025 submission
% index{Feng, Zetian}
% index{Fu, Juan}
% index{Zou, Xuebin}
% index{Ye, Hongsheng}
% index{Wu, Hong}
% index{Zhou, Jianhua}
% index{Wang, Yi}

\authorrunning{Z. Feng et al.}
\institute{
Smart Medical Imaging, Learning and Engineering (SMILE) Lab,
Medical UltraSound Image Computing (MUSIC) Lab,
School of Biomedical Engineering,
Shenzhen University Medical School,
Shenzhen University, Shenzhen, China \and
The Department of Ultrasound, State Key Laboratory of Oncology in South China,
Guangdong Provincial Clinical Research Center for Cancer,
Sun Yat-sen University Cancer Center, Guangzhou, China\\
\email{zhoujh@sysucc.org.cn}, \email{onewang@szu.edu.cn}
}

\maketitle              % typeset the header of the contribution
\begin{abstract}
Prostate cancer (PCa) is a leading cause of cancer-related mortality in men, and accurate identification of clinically significant PCa (csPCa) is critical for timely intervention.
Transrectal ultrasound (TRUS) is widely used for prostate biopsy; however, its low contrast and anisotropic spatial resolution pose diagnostic challenges. 
To address these limitations, we propose a novel hybrid-view attention (HVA) network for csPCa classification in 3D TRUS that leverages complementary information from transverse and sagittal views.
Our approach integrates a CNN-transformer hybrid architecture, where convolutional layers extract fine-grained local features and transformer-based HVA models global dependencies.
Specifically, the HVA comprises intra-view attention to refine features within a single view and cross-view attention to incorporate complementary information across views.
Furthermore, a hybrid-view adaptive fusion module dynamically aggregates features along both channel and spatial dimensions, enhancing the overall representation.
Experiments are conducted on an in-house dataset containing 590 subjects who underwent prostate biopsy.
Comparative and ablation results prove the efficacy of our method.
\textit{The code is available at}
\url{https://github.com/mock1ngbrd/HVAN}.

\keywords{Clinically significant prostate cancer \and Transrectal ultrasound \and Attention mechanism \and Computer-aided diagnosis.}
% Authors must provide keywords and are not allowed to remove this Keyword section.

\end{abstract}
\section{Introduction}
Prostate cancer (PCa) is one of the most common malignancies worldwide and the second leading cause of cancer-related deaths in men~\cite{siegel2025cancer}.
Prostate biopsy is the golden standard for diagnosing PCa.
Based on pathological features of prostate tissue, the diagnosis is classified into clinically significant PCa (csPCa), clinically insignificant PCa (cisPCa), and benign prostate hyperplasia~\cite{matoso2019defining}.
Identifying csPCa is crucial for timely intervention, as it has a poor prognosis and requires immediate treatment~\cite{mottet2021eau}.

Multiparametric magnetic resonance imaging (mp-MRI) is recommended for prostate biopsy, offering detailed localization of PCa~\cite{mottet2021eau}.
However, its limited availability and complicated operation hinder its widespread adoption.
Transrectal ultrasound (TRUS), with its low cost, ease of use, and real-time imaging capabilities, is commonly used for biopsy guidance~\cite{grey2022multiparametric}.
Yet, the low contrast of TRUS poses challenges in accurately identifying and diagnosing PCa~\cite{correas2021advanced}, leading to unnecessary biopsies and increased surgical risks~\cite{loeb2013systematic}.
Thus, methods to enhance early detection and accurate identification of csPCa in TRUS are critical.

Recently, deep learning approaches have been proposed for csPCa identification in TRUS.
Sun~\textit{et al}.~\cite{sun2023three} use a 3D convolutional neural network (CNN) with prostate mask guidance for csPCa classification in TRUS.
Wu~\textit{et al}.~\cite{wu2024multi} propose a multi-modality fusion network for csPCa classification, leveraging information from B-mode and shear wave elastography.
Later, Wu~\textit{et al}.~\cite{wu2024towards} extend their work by introducing few shot segmentation task to enhance the capability of classification encoder.
These methods succeed in identifying csPCa via additional guidance of prostate mask or elastic information.
However, they all focus on the transverse view, neglecting the complementary information in the sagittal view, which can help confirm any suspicious lesions~\cite{hendrikx2002trus}.
Therefore, it highlights the need for incorporating transverse-sagittal-view information to improve csPCa identification.
Combining information from different views has been explored in other medical imaging contexts~\cite{pi2020automated,song2022judgment,huang2022personalized},
but most approaches are limited to 2D space thus cannot be directly employed to analyze TRUS scan videos from both transverse and sagittal views.

In this study, we propose a hybrid-view attention (HVA) network for the accurate csPCa classification in 3D TRUS.
The proposed HVA not only refines the features within each view, but also leverages cross-view information to compensate for information loss induced by the large slice thickness.
Additionally, we introduce a fusion module that adaptively aggregates features across both channel and spatial dimensions.
Comparative and ablation experiments on a large TRUS dataset with biopsy-proven PCa validate the efficacy of our method.
Our main contributions are summarized as follows:
\begin{itemize}
    \item[$\bullet$] To the best of our knowledge, this is the first study for the classification of csPCa by simultaneously analyzing both transverse and sagittal TRUS scans.
    To enable effective hybrid-view feature learning, we design a novel architecture that combines CNNs for local feature extraction and transformers for global feature modeling.
    
    \item[$\bullet$] We introduce a hybrid-view attention, comprising intra-view attention (IVA) and cross-view attention (CVA).
    The IVA focuses on capturing fine-grained features within a single view, while the CVA integrates complementary information from orthogonal views.
    
    \item[$\bullet$] To effectively aggregate hybrid-view information, a dynamic feature fusion module is proposed to adaptively re-weight features along both channel and spatial dimensions.
\end{itemize}

\section{Method}

\begin{figure*}[t]
	\centering
	\includegraphics[width=\textwidth]{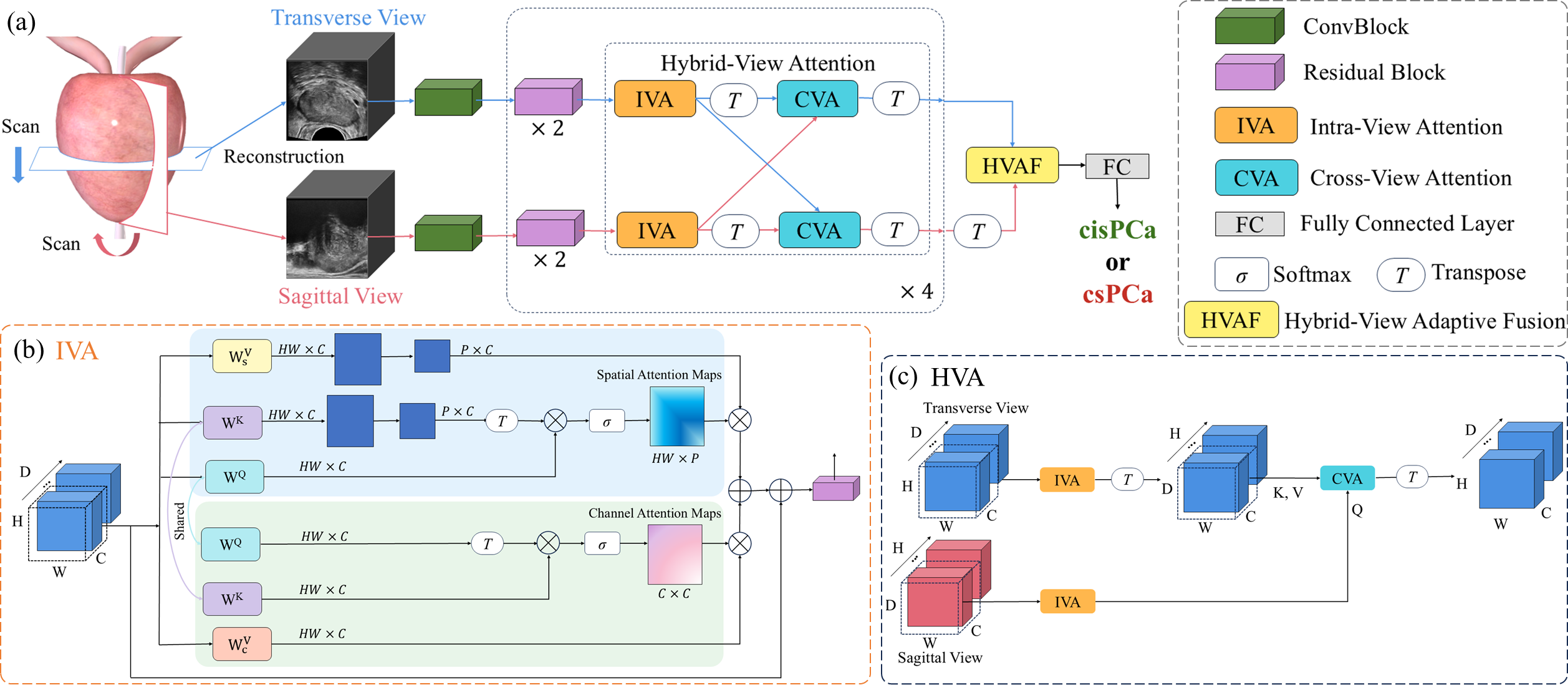}
	\caption{Overview of the proposed network: (a) network architecture, (b) intra-view attention module, and (c) the detailed structure in the hybrid-view attention module.}
	\label{fig: framework}
\end{figure*}

Fig.~\ref{fig: framework}(a) illustrates the architecture of our hybrid-view attention network, which employs a dual-stream encoder to extract and aggregate features from different views for classification.
The network leverages a CNN-transformer hybrid design, where each encoder is organized into four stages.
At each stage, two convolutional blocks first capture local features, followed by a HVA module that models global contextual information.
The HVA module comprises two specialized attention modules:
the IVA module refines features within a single view in the imaging plane,
while the CVA module leverages complementary information from an orthogonal view to further enhance feature quality.
Finally, a hybrid-view adaptive fusion (HVAF) module dynamically aggregates the refined features across both channel and spatial dimensions, yielding a comprehensive representation for accurate classification.

\subsection{CNN-Transformer Network for Hybrid-View Learning}
The proposed network employs a dual-stream encoder to capture view-specific features, facilitating intra- and inter-view feature learning.
Each stream consists of a stem layer followed by four down-sampling stages, each integrating a CNN-transformer hybrid design.

To establish local-to-global modeling, each stage comprises two residual convolutional blocks~\cite{he2016deep}, IVA and CVA modules, successively.
The two attention modules leverage attention mechanisms in transformer~\cite{vaswani2017attention} for long-range dependency modeling.
Specifically, the residual blocks extract fine-grained spatial details and local structures,
while the IVA module enhances features within a single view by capturing long-range dependencies.
The following CVA module further refines features by incorporating complementary spatial information from the orthogonal view.
Together, these two modules form the HVA module, modeling global features across different TRUS scan views.

By combining CNN-based local feature extraction with attention-driven global modeling,
the proposed network effectively captures both fine-grained details and long-range dependencies,
leveraging complementary features from different views for the csPCa classification task. 

\subsection{Hybrid-View Attention}
To model global dependencies across TRUS scans from different views,
we introduce a hybrid-view attention transformer module, which integrates both intra-view and cross-view attention modules.
As shown in Fig.~\ref{fig: framework}, the IVA module refines features within a single view, while the CVA module leverages high-quality spatial information from an orthogonal view to enhance feature representation.

\subsubsection{Intra-View Attention}
To fully leverage the high in-plane spatial resolution of TRUS images,
we propose IVA module to capture feature representations within the imaging plane of a specific view,
capturing fine-grained spatial dependencies crucial for the subsequent cross-view learning.

Given transverse and sagittal feature maps \(F_t, F_s \in \mathbb{R}^{B \times C \times H \times W \times D}\),
where \(B\) is the batch size, \(C\) is the number of channels, and \(H \times W \times D\) are spatial dimensions,
note that the transverse plane corresponds to \(H \times W\), while the sagittal plane corresponds to \(W \times D\).
To capture fine-grained information within the imaging view,
the imaging axis is merged into the batch dimension,
resulting in transverse features \(\tilde{F}_t \in \mathbb{R}^{BD \times C \times H \times W}\) and sagittal features \(\tilde{F}_s \in \mathbb{R}^{BH \times C \times W \times D}\),
on which IVA is applied.
Inspired by advanced attention mechanisms for contextual spatial-channel feature aggregation,
the IVA module is built upon an efficient paired-attention block~\cite{shaker2024unetr++}.
Each block consists of spatial and channel attention modules, designed to jointly capture spatial relationships and channel-wise dependencies.

Specifically, the IVA module employs a shared queries-keys mechanism to improve computational efficiency while preserving representational richness.
Taking the transverse view as an example, the input features \(\tilde{F}_t\) are processed in parallel by the spatial and channel attention modules, as shown in Fig.~\ref{fig: framework}(b). 
A shared set of query and key embeddings is learned, while separate value embeddings are used for spatial and channel attention.
Given \( \tilde{F}_t \in \mathbb{R}^{BD \times C \times H \times W} \),
we first project \( \tilde{F}_t \) into query \( Q_{shared} \), key \( K_{shared} \), spatial value \( V_{spatial} \), and channel value \( V_{channel} \) using four distinct linear layers:
\[
Q_{shared} = W^Q \tilde{F}_t, \quad K_{shared} = W^K \tilde{F}_t, \quad V_{spatial} = W^V_s \tilde{F}_t, \quad V_{channel} = W^V_c \tilde{F}_t.
\]
These projections have dimensions \( HW \times C \).
To reduce the computational burden of spatial attention, 
\(K_{shared}\) and \(V_{spatial}\) are projected into a lower-dimensional space using learnable matrices,
from \(HW \times C\) to \(P \times C\), where \(P \ll HW\).
The spatial attention is then computed as:
\begin{equation}
\label{eq: sa of iva}
F_{spatial} = softmax(\frac{Q_{shared} K_{proj}^T}{\gamma_s}) \cdot V_{proj},
\end{equation}
where \(Q_{shared}\), \(K_{proj}\), and \(V_{proj}\) denote shared queries, projected shared keys, projected spatial values, and \(\gamma_s\) is a learnable scaling parameter.
The channel attention captures inter-dependencies between feature channels by applying a matrix multiplication operation in the channel dimension between the channel values and channel attention maps. 
The channel attention is defined as follows:
\begin{equation}
\label{eq: ca of iva}
F_{channel} =  V_{channel} \cdot softmax(\frac{Q_{shared}^T K_{shared}}{\gamma_c}),
\end{equation}
where \(Q_{shared}\), \(K_{shared}\), and \(V_{channel}\) denote shared queries, shared keys,  channel values, and \(\gamma_c\) is a learnable scaling parameter.
Finally, the outputs of the spatial and channel attention modules are summed, together with the \(\tilde{F}_t\).
The aggregated features are reshaped back to \(\mathbb{R}^{B \times C \times H \times W \times D} \),
and further refined using a residual block to enhance representations.

\subsubsection{Cross-View Attention}
The slice thickness results in lower spatial resolution along the scanning direction comparing to the in-plane space.
While interpolation can address this imbalance, it does not provide any information gain and may introduce additional errors.
To address this issue, we propose a CVA module, which incorporates information from the orthogonal view to enhance feature representation.

The CVA module operates similarly to the IVA module, but instead of refining features in a single view, it applies attention across the orthogonal views.
Features from one view serve as queries to refine the features of the other view, capturing high-quality complementary orthogonal information to enrich the feature maps and improve performance in cross-view learning.
Specifically, the CVA module utilizes parallel spatial and channel attention modules.
Taking transverse view as an example (see Fig.~\ref{fig: framework}(c)), CVA is applied on the sagittal plane, using sagittal view features as high-quality queries, and vice versa.
Given feature maps \(F_t, F_s \in \mathbb{R}^{B \times C \times H \times W \times D}\),
the process begins by reshaping transverse and sagittal view features into 2D sagittal planar features, denoted as \(\tilde{F}_{ts} \in \mathbb{R}^{BH \times C \times W \times D}\) and \(\tilde{F}_s \in \mathbb{R}^{BH \times C \times W \times D}\), respectively. 
\(K_{shared}\), \(V_{spatial}\) and \(V_{channel}\) are projected from \(\tilde{F}_{ts}\), and \(Q_{shared}\) is projected from \(\tilde{F}_s\).
Finally, we apply the same computations as in IVA to refine the features, following Equations \ref{eq: sa of iva} and \ref{eq: ca of iva}.
After computing the HVA, the feature maps are reshaped back to \(\mathbb{R}^{B \times C \times H \times W \times D}\).

\subsection{Hybrid-View Adaptive Fusion}
\begin{figure*}[t]
	\centering
	\includegraphics[width=0.9\textwidth]{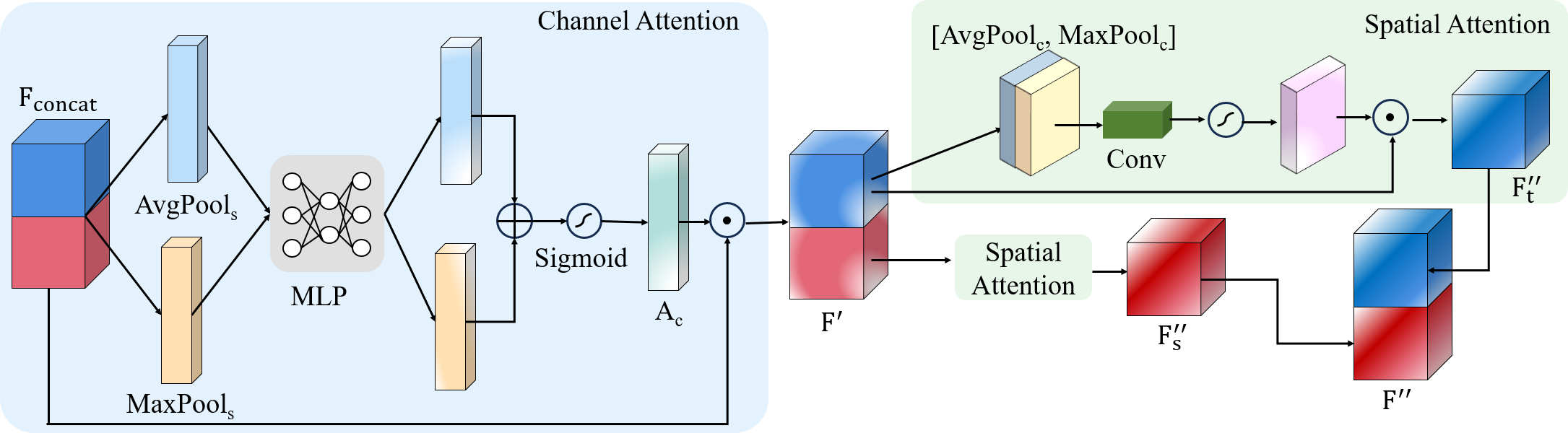}
	\caption{The structure of the proposed hybrid-view adaptive fusion module.}
	\label{fig: mvaf}
\end{figure*}

We propose the HVAF module to enhance the feature aggregation from different views by dynamically re-weighting features across both channel and spatial dimensions.
As illustrated in Fig.~\ref{fig: mvaf},
the HVAF module first concatenates the feature maps of both views, then applies channel attention to emphasize discriminative channels, followed by spatial attention to refine location-specific details within each view.
The final refined features are concatenated to produce the fusion output.

As shown in Fig.~\ref{fig: mvaf}, the channel attention mechanism in HVAF captures informative features across channels using both global average pooling and max pooling.
A gating mechanism based on multi-layer perceptron (MLP) adaptively re-weights feature maps. 
Given feature maps \(F_t, F_s \in \mathbb{R}^{C \times H \times W \times D}\),
the channel attention process is formulated as follows:
\begin{equation}
	\label{eq:1}
	F_{concat} = Concat(F_{t}, F_{s}),
\end{equation}
\begin{equation}
	\label{eq:2}
	F' = A_c(F_{concat}) \odot F_{concat},
\end{equation}
where \(Concat\) is the concatenation operation along the channel dimension,
\(\odot\) denotes element-wise multiplication,
and \(A_c \in \mathbb{R}^{C \times 1 \times 1 \times 1}\) is the channel attention map calculated as:
\begin{equation}
	\label{eq: ca of fusion}
	A_{c} = \sigma(MLP(AvgPool_s(F_{concat})) + MLP(MaxPool_s(F_{concat}))),
\end{equation}
where \(\sigma\) is sigmoid function, \(MLP\) is a shared multilayer perceptron, \(AvgPool_s\) and \(MaxPool_s\) are global average pooling and max pooling along spatial dimensions.
Note that the attention maps are broadcasted accordingly when multiplying, that is to say the channel attention map is broadcasted across the spatial dimensions and vice versa.

The spatial attention mechanism in HVAF is applied to each view independently.
It highlights salient spatial regions by combining both average and max pooling across channels, enabling the model to focus on important anatomical semantic information.
The features of different views are separated from $F'$:
\begin{equation}
	\label{eq:3}
	F'_{t} = F'[:C], F'_{s} = F'[C:],
\end{equation}
then the spatial attention is computed as follows:
\begin{equation}
	\label{eq:4}
	F''_{t} = A_s(F'_{t}) \odot F'_{t}, F''_{s} = A_s(F'_{s}) \odot F'_{s},
\end{equation}
\begin{equation}
	\label{eq:5}
	F'' = Concat(F''_{t}, F''_{s}),
\end{equation}
where $F''\in \mathbb{R}^{C \times H \times W \times D}$ is the final fused features, 
\(A_s \in \mathbb{R}^{1 \times H \times W \times D}\) is the spatial attention map calculated as:
\begin{equation}
\label{eq: sa of fusion}
A_{s} = \sigma(Conv(Concat(AvgPool_c(F), MaxPool_c(F)))),
\end{equation}
where \(Conv\) is a \(3 \times 3 \times 3\) convolution,
\(AvgPool_c\) and \(MaxPool_c\) are global average pooling and max pooling along channel dimensions.

By integrating channel and spatial attentions, the HVAF module adaptively fuses features from different views, enabling the network to aggregate complementary information for improved classification performance.

\begin{table}[t]
	\centering
	\renewcommand{\tabcolsep}{1.2mm}                              
	\caption{Comparative and ablation results of different methods (best results are highlighted in bold).}
	\label{tab: result}
	\small
	\begin{tabular}{cccccccccc}
		\toprule
		TV     & SV        & IVA     &CVA    &HVAF                             & AUC      & F1-score       & Accuracy &Sensitivity  &Specificity\\
		\midrule
		\checkmark &       &        &       &                                  & 0.7562     & 0.8000     & 0.7171  &0.8269   &0.4792\\
            \checkmark &       &\checkmark&     &                                    & 0.7808     & 0.823     & 0.7566  &0.8269   &0.6042\\
		       &\checkmark   &            &          &                            & 0.7015     & 0.7729     & 0.6908  &0.7692   &0.5208\\
             &    \checkmark   &\checkmark&     &                                    & 0.7600     & 0.8246     & 0.7368  &0.9038   &0.3750\\
		\checkmark & \checkmark &            &          &                        & 0.7622     & 0.8241     & 0.7500  &0.8558   &0.5208\\
		\checkmark & \checkmark & \checkmark &          &                        & 0.8149     & 0.8235     &0.7237   &\textbf{0.9423}   &0.2500\\ 
		\checkmark & \checkmark & \checkmark & \checkmark  &                     & 0.8249     & 0.8042     &0.7566   &0.7308   &\textbf{0.8125}\\
        \checkmark & \checkmark & \checkmark & \checkmark  & \checkmark           
		& \textbf{0.8534}     & \textbf{0.8611}         &\textbf{0.8026}   & 0.8942 & 0.6042\\   
		\midrule
        \multicolumn{5}{c}{DMVFN~\cite{song2022judgment}}                        & 0.7622 & 0.7795
        & 0.7171 & 0.7308 & 0.6875 \\
        \multicolumn{5}{c}{AADNN~\cite{pi2020automated}}                        & 0.7893 & 0.7897
        & 0.7303 & 0.7404 & 0.7083 \\
        \multicolumn{5}{c}{MVMT~\cite{huang2022personalized}}                   & 0.7843 & 0.8295
        & 0.7566 & 0.8653 & 0.5208 \\
        \multicolumn{5}{c}{MMFN~\cite{wu2024multi}}          & 0.7812 & 0.8225
        & 0.7303 & 0.9135 & 0.3333 \\
		\bottomrule
	\end{tabular}
\end{table}

\section{Experiments}
\subsubsection{Dataset}
The experiments were carried out on an in-house 3D TRUS dataset collected from the Cancer Center of Sun Yat-Sen University.
The study was conducted retrospectively and therefore receiving a waiver of approval from the institutional review board.
The dataset comprises 590 pairs of TRUS volumes scanned from transverse and sagittal views,
collected from 590 patients who underwent TRUS-guided transperineal biopsy followed by whole-gland prostatectomy.
It includes 401 csPCa and 189 cisPCa cases.
We randomly split the dataset into training set (438 pairs, 297 csPCa) and testing set (152 pairs, 104 csPCa).
All volumes were resized to \(128 \times 128 \times 128\).

\subsubsection{Implementation Details}
The method was implemented in PyTorch on a NVIDIA Tesla V100 GPU with 32G memory.
We trained the network using Adam optimizer with an initial learning rate of \(10 ^ {-4}\) for 100 epochs.
Focal loss~\cite{lin2017focal} was employed.
The following quantitative metrics were used to evaluate the classification performance~\cite{WANG2025189}:
area under the ROC curve (AUC), F1-score (F1), accuracy, sensitivity, and specificity.

\subsubsection{Results}
\begin{figure*}[t]
	\centering
	\includegraphics[width=0.95\textwidth]{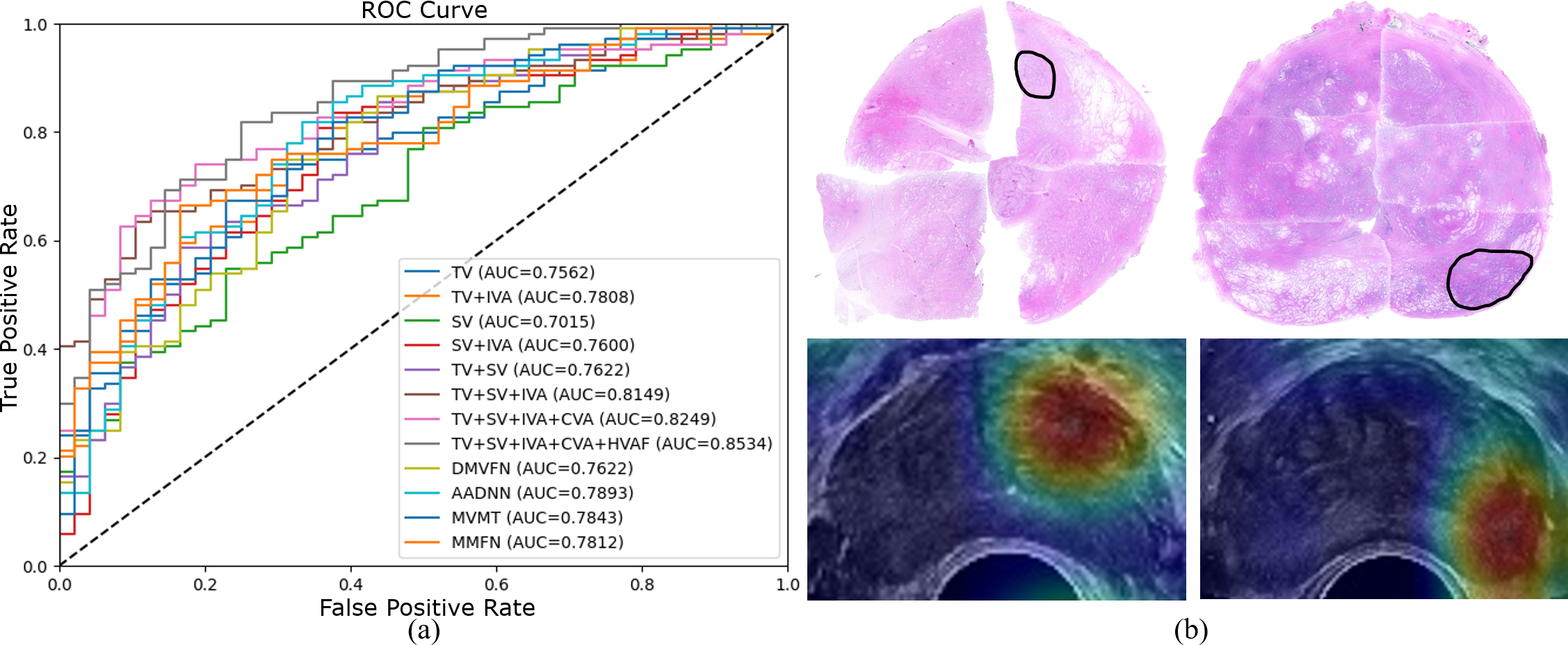}
	\caption{(a) The receiver operating characteristic (ROC) curves of different methods on the testing set. (b) The class activation mapping (CAM) images generated using our network, and the pathological images with annotated csPCa locations.}
	\label{fig: roc}
\end{figure*}

We first assessed the contribution of each designed component through ablation experiments.
Firstly, we evaluated single-view performance using only one view (TV or SV) and one stream of encoder.
Secondly, we analyzed the impact of the IVA module in both single-view and multi-view settings.
Thirdly, we examined the effect of the CVA module.
Finally, we accessed the HVAF module by replacing it with simple feature concatenation.
Table~\ref{tab: result} shows the ablation results.
Fig.~\ref{fig: roc}(a) presents the ROC curves.
It can be observed that simply concatenating features from both views improved AUC compared to single-view model.
This highlights the complementary nature of multi-view images.
The addition of the IVA module enhanced the model performance in both single-view and multi-view settings, while the CVA module further improved AUC and accuracy of the model.
Notably, the HVAF module led to substantial gains in AUC, F1-score, and accuracy.

We further compared our method with cutting-edge multi-view fusion approaches~\cite{song2022judgment,pi2020automated,huang2022personalized}.
Since these approaches are applied in other 2D medical imaging contexts, we adapted them to 3D by replacing 2D convolution and other 2D operations with their 3D counterparts.
Additionally, we compared a related multi-modal fusion model for csPCa classification~\cite{wu2024multi}.
As shown in Table~\ref{tab: result},
our method consistently outperformed these approaches in terms of AUC, F1-score and accuracy,
underscoring its efficacy in identifying csPCa in 3D TRUS through our hybrid-view learning method.

Fig.~\ref{fig: roc}(b) visualizes two examples of 2D slices with class activation mapping (CAM) overlaid.
They are extracted from two csPCa TRUS volumes, respectively.
Fig.~\ref{fig: roc}(b) also shows their corresponding pathological images with annotated csPCa locations.
It can be observed a favorable agreement of the suggested csPCa regions.

\section{Conclusion}
We propose a hybrid-view attention network for csPCa classification in 3D TRUS.
The hybrid-view attention comprises intra-view attention to refine features within a single view and cross-view attention to leverage complementary information across orthogonal views.
In addition, a hybrid-view adaptive fusion module dynamically aggregates features along both channel and spatial dimensions. 
Extensive experiments on an in-house TRUS dataset demonstrate that our approach effectively outperforms single-view methods and state-of-the-art multi-view techniques,
underscoring the capability of our hybrid-view learning in improving prostate cancer diagnosis.
%Future work is to validate our method on a large multi-center dataset to evaluate its generalization performance.

\begin{credits}
	\subsubsection{\ackname}
	This work was supported
	in part by the National Natural Science Foundation of China under Grants 81971631, 8320108011, 62471306 and 62071305,
	in part by the Shenzhen Medical Research Fund under Grant D2402010,
	in part by the Guangdong-Hong Kong Joint Funding for Technology and Innovation under Grant 2023A0505010021,
	and in part by the Guangdong Basic and Applied Basic Research Foundation under Grant 2022A1515011241.\\
	\\
	This preprint has not undergone peer review or any post-submission improvements or corrections. The Version of Record of this contribution is published in ***, and is available online at ***.
	
	\subsubsection{\discintname}
	The authors have no competing interests to declare that are relevant to the content of this article.
\end{credits}
%
% ---- Bibliography ----

\bibliographystyle{splncs04}
\bibliography{Paper-1144}

\begin{thebibliography}{10}
\providecommand{\url}[1]{\texttt{#1}}
\providecommand{\urlprefix}{URL }
\providecommand{\doi}[1]{https://doi.org/#1}

\bibitem{correas2021advanced}
Correas, J.M., Halpern, E.J., Barr, R.G., Ghai, S., Walz, J., Bodard, S.,
  Dariane, C., de~la Rosette, J.: Advanced ultrasound in the diagnosis of
  prostate cancer. World Journal of Urology  \textbf{39},  661--676 (2021)

\bibitem{grey2022multiparametric}
Grey, A.D., Scott, R., Shah, B., Acher, P., Liyanage, S., Pavlou, M., Omar, R.,
  Chinegwundoh, F., Patki, P., Shah, T.T., et~al.: Multiparametric ultrasound
  versus multiparametric {MRI} to diagnose prostate cancer ({CADMUS}): a
  prospective, multicentre, paired-cohort, confirmatory study. The Lancet
  Oncology  \textbf{23}(3),  428--438 (2022)

\bibitem{he2016deep}
He, K., Zhang, X., Ren, S., Sun, J.: Deep residual learning for image
  recognition. In: Proceedings of the IEEE Conference on Computer Vision and
  Pattern Recognition (CVPR). pp. 770--778 (2016)

\bibitem{hendrikx2002trus}
Hendrikx, A., Safarik, L., Hammerer, P.: {TRUS} and biopsy: practical aspects.
  European Urology  \textbf{41}(6), ~I--X (2002)

\bibitem{huang2022personalized}
Huang, H., Dong, Y., Jia, X., Zhou, J., Ni, D., Cheng, J., Huang, R.:
  Personalized diagnostic tool for thyroid cancer classification using
  multi-view ultrasound. In: International Conference on Medical Image
  Computing and Computer-Assisted Intervention (MICCAI). pp. 665--674. Springer
  (2022)

\bibitem{lin2017focal}
Lin, T.Y., Goyal, P., Girshick, R., He, K., Doll{\'a}r, P.: Focal loss for
  dense object detection. In: Proceedings of the IEEE International Conference
  on Computer Vision. pp. 2980--2988 (2017)

\bibitem{loeb2013systematic}
Loeb, S., Vellekoop, A., Ahmed, H.U., Catto, J., Emberton, M., Nam, R.,
  Rosario, D.J., Scattoni, V., Lotan, Y.: Systematic review of complications of
  prostate biopsy. European Urology  \textbf{64}(6),  876--892 (2013)

\bibitem{matoso2019defining}
Matoso, A., Epstein, J.I.: Defining clinically significant prostate cancer on
  the basis of pathological findings. Histopathology  \textbf{74}(1),  135--145
  (2019)

\bibitem{mottet2021eau}
Mottet, N., van~den Bergh, R.C., Briers, E., Van~den Broeck, T., Cumberbatch,
  M.G., De~Santis, M., Fanti, S., Fossati, N., Gandaglia, G., Gillessen, S.,
  et~al.: {EAU-EANM-ESTRO-ESUR-SIOG} guidelines on prostate cancer—2020
  update. {P}art 1: screening, diagnosis, and local treatment with curative
  intent. European Urology  \textbf{79}(2),  243--262 (2021)

\bibitem{pi2020automated}
Pi, Y., Zhao, Z., Xiang, Y., Li, Y., Cai, H., Yi, Z.: Automated diagnosis of
  bone metastasis based on multi-view bone scans using attention-augmented deep
  neural networks. Medical Image Analysis  \textbf{65},  101784 (2020)

\bibitem{shaker2024unetr++}
Shaker, A.M., Maaz, M., Rasheed, H., Khan, S., Yang, M.H., Khan, F.S.:
  {UNETR}++: delving into efficient and accurate 3{D} medical image
  segmentation. IEEE Transactions on Medical Imaging  \textbf{43}(9),
  3377--3390 (2024)

\bibitem{siegel2025cancer}
Siegel, R.L., Kratzer, T.B., Giaquinto, A.N., Sung, H., Jemal, A.: Cancer
  statistics, 2025. CA: A Cancer Journal for Clinicians  (2025)

\bibitem{song2022judgment}
Song, D., Zhang, Z., Li, W., Yuan, L., Zhang, W.: Judgment of benign and early
  malignant colorectal tumors from ultrasound images with deep multi-view
  fusion. Computer Methods and Programs in Biomedicine  \textbf{215},  106634
  (2022)

\bibitem{sun2023three}
Sun, Y.K., Zhou, B.Y., Miao, Y., Shi, Y.L., Xu, S.H., Wu, D.M., Zhang, L., Xu,
  G., Wu, T.F., Wang, L.F., et~al.: Three-dimensional convolutional neural
  network model to identify clinically significant prostate cancer in
  transrectal ultrasound videos: a prospective, multi-institutional, diagnostic
  study. EClinicalMedicine  \textbf{60} (2023)

\bibitem{vaswani2017attention}
Vaswani, A., Shazeer, N., Parmar, N., Uszkoreit, J., Jones, L., Gomez, A.N.,
  Kaiser, {\L}., Polosukhin, I.: Attention is all you need. Advances in Neural
  Information Processing Systems  \textbf{30} (2017)

\bibitem{WANG2025189}
Wang, H., Wu, H., Wang, Z., Yue, P., Ni, D., Heng, P.A., Wang, Y.: A narrative
  review of image processing techniques related to prostate ultrasound.
  Ultrasound in Medicine \& Biology  \textbf{51}(2),  189--209 (2025)

\bibitem{wu2024multi}
Wu, H., Fu, J., Ye, H., Zhong, Y., Zou, X., Zhou, J., Wang, Y.: Multi-modality
  transrectal ultrasound video classification for identification of clinically
  significant prostate cancer. In: 2024 IEEE International Symposium on
  Biomedical Imaging (ISBI). pp.~1--5. IEEE (2024)

\bibitem{wu2024towards}
Wu, H., Fu, J., Ye, H., Zhong, Y., Zou, X., Zhou, J., Wang, Y.: Towards
  multi-modality fusion and prototype-based feature refinement for clinically
  significant prostate cancer classification in transrectal ultrasound. In:
  International Conference on Medical Image Computing and Computer-Assisted
  Intervention (MICCAI). pp. 724--733. Springer (2024)

\end{thebibliography}
\end{document}